\documentclass[aip,graphicx, amsmath,amssymb,reprint]{revtex4-1}
\usepackage{graphicx}
\draft 
\usepackage{mhchem}
\usepackage{filecontents}
\usepackage{subcaption}
\usepackage{comment}
\usepackage{svg}
\usepackage{cleveref}

\begin{document}


\title{Ultra-low energy threshold engineering for all-optical switching of magnetization in dielectric-coated Co/Gd based synthetic-ferrimagnet} 



\author{Pingzhi Li}
\email[]{p.li1@tue.nl}
\author{Mark J.G. Peeters}
\email[]{m.j.g.peeters@tue.nl}
\author{Youri L.W. van Hees}
\email[]{y.l.w.v.hees@tue.nl}
\author{Reinoud Lavrijsen}
\email[]{r.lavrijsen@tue.nl}
\author{Bert Koopmans}
\email[]{b.koopmans@tue.nl}
\affiliation{Department of Applied Physics, Eindhoven University of Technology\\ 
P. O. Box 513, 5600 MB Eindhoven, The Netherlands}

\date{\today}

\begin{abstract}
A femtosecond laser pulse is able to switch the magnetic state of a 3d-4f ferrimagnetic material on a pico-second time scale.
Devices based on this all-optical switching (AOS) mechanism are competitive candidates for ultrafast  memory applications. 
However, a large portion of the light energy is lost by reflection from the metal thin film as well as transmission to the substrate.
In this paper, we explore the use of dielectric coatings to increase the light absorption by the magnetic metal layer based on the principle of constructive interference. 
We experimentally show that the switching energy oscillates with the dielectric layer thickness following the light interference profile as obtained from theoretical calculations.
Furthermore, the switching threshold fluence can be reduced by at least $80\%$ to 0.6 mJ/cm$^2$ using two dielectric SiO$_2$ layers sandwiching the metal stack, which scales to 15 fJ of incident energy for a cell size of $50^2$ nm$^2$.  
\end{abstract}


\maketitle 

\section{Introduction}
All-optical switching (AOS) of magnetization by a femtosecond (fs) laser pulse\cite{Stanciu:2007aa},
discovered in 2007, is found to be the fastest way known-to-date to switch the magnetic state of spintronic materials \cite{Kimel_AOS_Review2019}. It is almost a thousand times faster than switching based on spin transfer torque (STT) and spin orbit torque (SOT)\cite{El-Ghazaly:2020aa}. It thus offers platforms to create ultrafast memories as well as bridging integration between spintronics and photonics\cite{Becker:2020aa,Lalieu:2019aa}.

Such a magnetic switching process was found notably in the 3d-4f ferrimagnetic material systems, such as GdFeCo alloys\cite{Stanciu:2007aa,Ostler:2012aa} and layered structures\cite{Lalieu:2017aa,Aviles-Felix:2019aa}. 
From these materials, the Co/Gd bilayer\cite{Lalieu:2017aa} has received considerable attention recently owing to its several advantages over other materials to be embedded in the integrated platforms.  Being a layered structure, the wafer scale production can be much easier realized. The occurrence of AOS does not demand a specific material composition\cite{Lalieu:2017aa,Beens:2019aa} like for the GdFeCo alloys, and Tb/Co multilayers\cite{Aviles-Felix:2020aa}. Moreover, the switching energy is much lower\cite{Lalieu:2017aa,Khorsand:2012aa} and it can withstand thermal annealing\cite{Wang:2020ab} required for the fabrication of opto-switchable magnetic tunnel junctions (MTJs)\cite{Chen:2017aa}, as requried and demonstrated for
an opto-MTJ with high tunnel-magnetic resistance based on Co/Gd\cite{wang2020picosecond}. In addition, the Co/Gd bilayer displays strong interface induced spintronic effects, such as perpendicular magnetic anisotropy, spin-hall effect and interfacial Dzyaloshinskii-Moriya interaction\cite{Blasing:2018aa}. 
Based on this notion, the Co/Gd bilayer system has a high potential to allow for a hybrid opto-spintronic racetrack memory platform based on AOS and domain wall motion\cite{Lalieu:2019aa}.

  Despite that AOS shows a lower energy footprint than the STT/SOT-MRAM\cite{Stanciu:2007aa,Kimel_AOS_Review2019,El-Ghazaly:2020aa}, the high power of the light pulse due to its short pulse requirement for AOS\cite{Gorchon:2016aa} might lead to significant non-linear absorption loss (such as two-photon absorption and subsequent free-carrier absorption)\cite{Latkowski:2015uw,Gonzalez:2009ut,Marculescu:2017vj,Sobolewska:2020aa} when used in photonic integrated circuits. Therefore, lowering the power, i.e. the required incident pulse energy per switch, will facilitate the realization of AOS in integrated photonics as well as making the energy footprint of AOS even more competitive than spintronic based switching. 
  
  In current work on AOS  in Co/Gd, the light absorption is far from efficient, as a significant portion of the incident light energy is lost by reflection from the metal surface and transmission to the substrate.  Such losses not only offset the advantages of AOS, but also impose potential danger to other established components in the photonic/electronic network. 
  In order to eliminate the losses from transmission and reflection, we propose to use dielectric coatings, which can tune the phase of light interacting with the metal thin film satisfying the condition of constructive interference, thus to optimize the light absorption by the metallic thin film. 
  
  In this report, we used $\text{SiO}_2$ as dielectric layers both on top and at the bottom of the Co/Gd bilayer to theoretically and experimentally demonstrate the application of this concept to AOS. We show that the threshold fluence of AOS can be lowered significantly with the help of adding a bottom and top dielectric layer. The threshold fluence oscillates with respect to the layer thickness, which corresponds to the light interference behaviour as obtained from simulations. Our result shows that by sandwiching the metal layer with two rather standard (SiO$_2$) dielectric layers with a thickness of quarter-wavelength, the threshold fluence can be reduced by more than 80\% as compared to an uncoated metal layer. 

  \begin{figure*}
  \centering
\includegraphics[scale=.32]{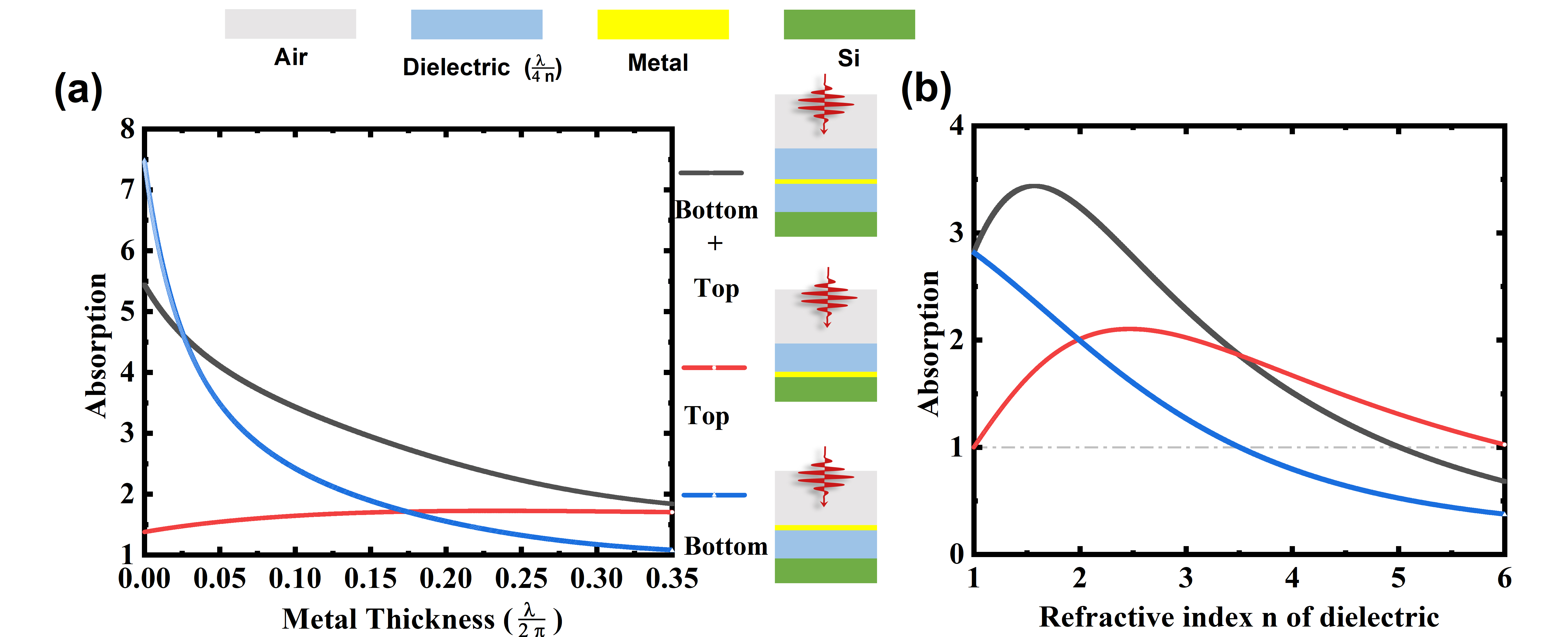}
\caption{Light absorption in a metal thin film normalized to the absorption without a dielectric coating, as a function of metal layer thickness assuming $n = 1.5$ (a) and of refractive index of dielectric material $n$ assuming the metal thickness is 0.1 $\frac{\lambda}{2\pi}$(b) , for three cases: only bottom $\frac{\lambda}{4 n}$ coating present (blue), only top $\frac{\lambda}{4 n}$ coating present (red) and both present (black).    
     }
\label{fig:anal}
\end{figure*}

\section{Theory}

Previous studies have revealed that AOS by a single linearly-polarized fs laser pulse is a heat driven process\cite{Radu:2011aa,Beens:2019aa,Kimel_AOS_Review2019}. The switching occurs at a sub-pico-second timescale due to ultrafast angular momentum exchange between two antiferromagnetically coupled sublattices \cite{Radu:2011aa}. In this paper, our theoretical analysis is based on the fact that metal absorbs light almost instantaneously enabled by electron excitations \cite{Kittel:2004aa}, and successive heat diffusion is small at a ps time scale. Furthermore, the threshold fluence for AOS is predominantly determined by the condition that the electron temperature should be raised above the Curie temperature of the magnetic film. Thus, we assume the threshold fluence to scale inversely proportionally with the local, instantaneous light absorption.  

In order to derive generic guidelines, we discuss the effect of adding dielectric layers to enhance the light absorption in (magnetic) metal thin film. The dielectric layers allow for multiple interactions (as a result of multiple reflections) of light with the metal film by introducing an extra interface. The optimum absorption occurs when the phase difference between each consecutive reflection to be (an odd number of) $\pi$, thus the thickness of the dielectric need to be an odd multiple of a quarter of the wavelength of the incoming light.  
Here, we adopted an analytical approach based on the transfer matrix method\cite{saleh2019fundamentals, byrnes2016multilayer} to qualitatively describe the influence of the metal thickness and refractive index $n$ of the dielectric material. As a representative example for typical 3d-4f bilayers, we assume a single metal layer with $n_m=4+4i$. We express the thickness of the metal in units of $\frac{\lambda}{2 \pi}$ for generic insights. In our calculation, we consider the absorption in the metal with a quarter-wavelength coating both on top and below the metal thin film normalized to that of the case without dielectrics as shown in Fig \ref{fig:anal}. When the thickness of the metal is below 0.03 $\frac{\lambda}{2\pi}$, a bottom coating only can already boost the absorption significantly (see Fig. \ref{fig:anal}a, $n$ = 1.5 assumed).
The bottom coating ensures optimized reflection from the Si substrate to be in phase to reduce the transmission loss. However, in this thickness range, the reflection loss from the top surface is negligible; adding a top coating additionally reduces the gain from the bottom coating due to its anti-reflection effect for light going from substrate to the air. As the metal thickness increases, the relative effect of the bottom layer decreases due to weakened transmission through the metal. 
The role of the top coating becomes gradually more significant due to increased reflection loss from the top metal surface. Especially, in the regime of a thick film ( $>\,$ 0.3 $\frac{\lambda}{2\pi}$), it is sufficient to have a top coating only. While in the thickness range within the skin depth of metal (0.03 $\frac{\lambda}{2\pi}$ $<$ t $<$ 0.2 $\frac{\lambda}{2\pi}$), combining a top with a bottom layer always leads to an overall gain. Knowing the relative role of the top and bottom coating, the $n$ of the dielectric is discussed next  (see Fig \ref{fig:anal}b.). For the top coating, the optimum $n$ for the anti-reflectivity is $\sqrt{|n_{m|}n_{Air}}\,\approx\,$ 2.2, while for the bottom coating, $n$ = 1 is desired. An optimum of $n$ occurs at 1.56 as top and bottom coatings work collaboratively (see Fig \ref{fig:anal}b.). Thus, SiO$_2$ ($n = 1.45$) has a value close to the optimum value.

\begin{figure}
    \centering
    \includegraphics[scale=.27]{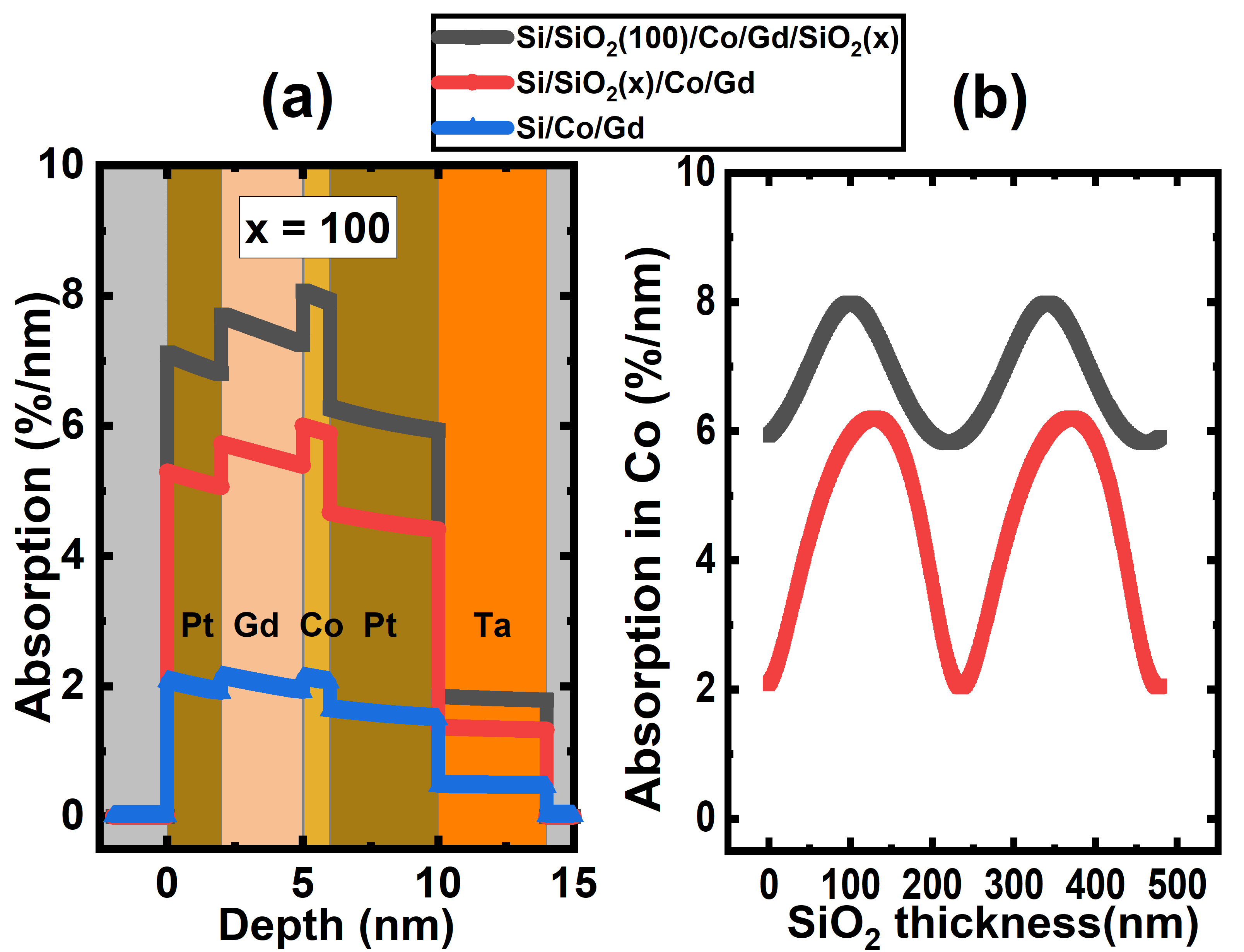}
    \caption{The depth profile of light absorption (wavelength 700 nm) calculated using the transfer matrix method.\cite{byrnes2016multilayer} (a) The light absorption profile for sample without dielectric coating (blue), 100 nm SiO$_2$ coating in the bottom(black), and 100 nm coating both at the bottom and top (black). (b) The light absorption in the Co layer as a function of the SiO$_2$ thickness for the Si/SiO$_2$($x$)/Co/Gd shown in red, and the Si/SiO$_2$(100)/Co/Gd/SiO$_2$($x$) shown in black. }
    \label{fig:sim}
\end{figure}

We proceeded by performing a calculation of light absorption in a specific metallic Co/Gd layer stack 
using $\lambda$ = 700 nm, where
the refractive index of each layer was taken from Palik \textit{et al.} \cite{Edward-D.-Palik:1985aa}. The term Co/Gd is used in this paper to denote the full stack of Ta(4)/Pt(4)/Co(1)/Gd(3)/Pt(2) from bottom to top (thickness in nm in parentheses) on a Si substrate. In the calculation, we also introduced one or two SiO$_2$ layers. In Fig. 2a, we show the depth profile of the light absorption by Si/Co/Gd first without any SiO$_2$ layer (see blue line), which in total accounts for about 20\%  of total light absorption. With the help of a bottom layer of SiO$_2$(100) between Co/Gd and Si, the light absorption in Co/Gd can be enhanced by a factor $\sim$ three (see red line in Fig. \ref{fig:sim}a). The enhancement can be extended further reaching a gain of 4, by adding another top layer of SiO$_2$(100) (see black line in Fig. 2a), which confirms the former generic discussion as the thickness of Co/Gd corresponds to 0.08 $\frac{\lambda}{2\pi}$. 
The enhancement is enabled by the constructive interference of light interacting with the Co/Gd. This phenomenon can be further demonstrated by showing the results in Fig 2b, in which upon varying the thickness of the SiO$_2$ layer, the light absorption shows oscillatory behaviour, where the extrema are occuring at an odd multiple of a quarter wavelength. Our results show that the light absorption in Co/Gd is expected to be enhanced significantly by applying two SiO$_2$ coatings sandwiching the Co/Gd. 
  
\begin{figure*}
\centering
\includegraphics[scale=0.3]{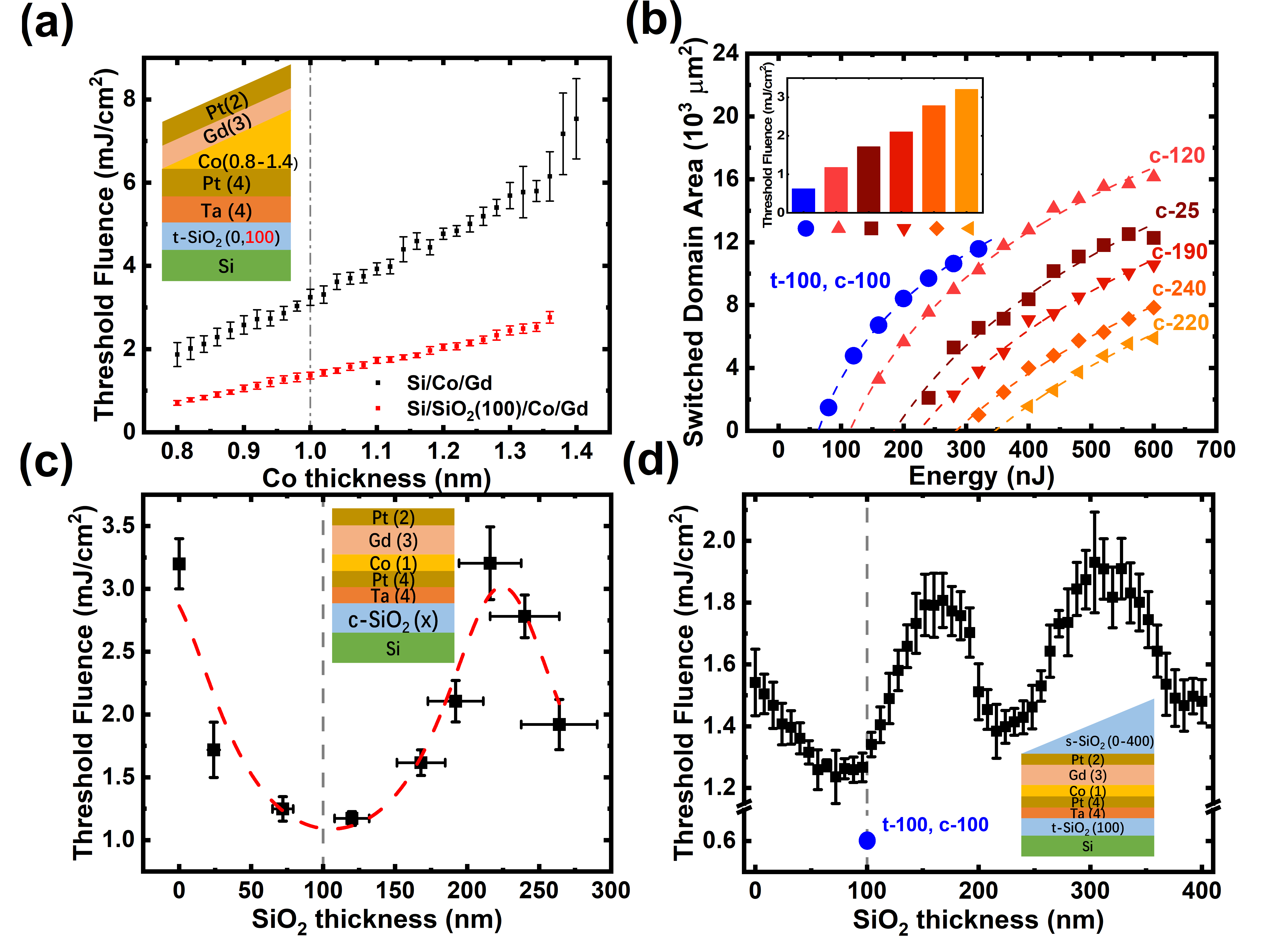}
\caption{(a) The threshold fluence of Co($x$)/Gd as a function of Co thickness on Si (black) and Si/t-SiO$_2$(100) (red) substrate.  
    (b) The switched domain size as a function of pulse energy for various samples with different ICP-CVD SiO$_2$ layer thickness, which is marked as a number (unit nm) next to each corresponding curve. All curves are for samples without top-dielectric, except for the blue data set, which is for Co/Gd with both a 100 nm thermal  SiO$_2$ (t-SiO$_2$) underlayer and a 100 nm ICV-CVD  SiO$_2$ (c-SiO$_2$) top layer. The threshold fluence for each curve obtained by the method from Ref.\cite{Lalieu:2017aa} is displayed in the inset. 
    (c) The threshold fluence of Co/Gd as a function of the thickness of ICP-CVD SiO$_2$. The red dashed line is the fitting obtained by taking the reciprocal of the sinusoidal oscillation as shown in Fig. \ref{fig:sim}b. 
    (d)  The threshold fluence as a function of the thickness of the top SiO$_2$ for Co/Gd grown on Si/SiO$_2$. 
     }
\label{fig:fig2}
\end{figure*}
\section{Methodology}
In this paper, we experimentally studied the influence of SiO$_2$ on the threshold fluence of AOS in Co/Gd following our calculation shown in Fig. \ref{fig:sim}. We begin with discussing the method for sample fabrication, which is followed by introducing the experimental method to determine the threshold fluence of AOS.

  The magnetic Co/Gd bilayer stacks measured in this study were deposited using DC magnetron sputtering in a chamber with a base pressure of $5\times 10^{-9}$ mBar. The Co/Gd exhibits perpendicular magnetic anisotropy with 100\% remanence due to the induced anisotropy at the Pt/Co interface. The magnetic properties of Co/Gd are described in Ref.\cite{Lalieu:2017aa}. We created several samples to experimentally demonstrate the concepts presented in Fig. \ref{fig:sim}. A first set of samples, termed Si/Co(x)/Gd without SiO2, and Si/t-SiO$_2$(100)/Co(x)/Gd with a 100 nm of thermally oxidized SiO$_2$ (t-SiO$_2$) coating is fabricated by wedge-sputtering Co(x)/Gd (a moving wedge shutter incorporated during sputtering causing the thickness of deposited material varying with position) both on a Si substrate and a Si substrate with a 100 nm thermal SiO$_2$ coating (t-SiO$_2$) as a single batch. The sample created in this way provides us with a very convenient way to study the thickness dependence without the needs of creating multiple samples. The schematic illustration of this set of samples is shown as the inset in Fig. \ref{fig:fig2}a.  A second set of sample: as Si/c-SiO$_2$(x)/Co/Gd, is made by sputtering Co/Gd on several Si wafer blankets pre-coated with different SiO$_2$ by inductively-coupled plasma chemical-vapour deposition (ICP-CVD) (c-SiO$_2$).  The c-SiO$_2$ in this sample set is deposited at 80 $^{\circ}$C, which preserves an atomically flat surface for the growth of the Co/Gd. The illustration of this set of samples is shown in the inset of Fig. \ref{fig:fig2}c. A third set of samples, termed as Si/t-SiO$_2$(100)/Co/Gd/s-SiO$_2$(x), is created by wedge sputtering a SiO$_2$ layer (s-SiO$_2$) on top of Si/t-SiO$_2$/Co/Gd.
  However, its properties deviate from stoichiometric SiO$_2$ (such as t-SiO$_2$ and c-SiO$_2$), its refractive index is higher 
  (due to deficiencies of O) provided with a small imaginary part of refractive index \cite{Hacker:1982aa,Khodier:2001aa}. The deposition procedure also causes damage to Co/Gd as we will discuss later. A last sample, termed as Si/t-SiO$_2$(100)/Co/Gd/c-SiO$_2$(100), is created, the Co/Gd of which is deposited from the same batch of Si/c-SiO$_2$(x)/Co/Gd followed by being coated with a top c-SiO$_2$(100).

 To determine the AOS threshold fluence, we adopted the method used by Lalieu $et$ $al.$ \cite{Lalieu:2017aa} as briefly described in the following. Samples are firstly saturated with a magnetic field in the out-of-plane direction.  We then created domains by AOS, by illuminating the magnetic thin film normal to the sample plane by a single linearly polarized fs laser pulse (center wavelength 700 nm, pulse duration 50 fs) with varying energy. 
A domain is created once the laser energy surpasses a certain energy threshold. The domain size increases upon further enhancing the energy,   the domain written has an Gaussian elliptical shape. We then image the switched domains by polar-Kerr microscopy and determine its area.  
The threshold fluence is obtained by fitting the switched domain area and the pulse energy, to a Gaussian elliptical spot profile. 
 Some examples of the measurement results together with the fitted curve can be found in Fig. \ref{fig:fig2}b, while the determined threshold fluence corresponding to each curve, can be found in the inset.

\section{Results and Discussions}

 We begin with discussing the effect of the SiO$_2$ bottom layer. In Fig. \ref{fig:fig2}a, the experimental results for Si/t-SiO$_2$(0,100)/Co($x$)/Gd is presented. It can be seen that the threshold fluence of Co($x$)/Gd with and without t-SiO$_2$(100) increases as a function of Co thickness as a result of increased Curie temperature \cite{Lalieu:2017aa,Beens:2019aa}.  
At each individual Co thickness, the threshold fluence of Co/Gd grown on Si is more than twice that of Co/Gd grown on Si/t-SiO$_2$(100), which confirms our calculation in Fig. \ref{fig:sim}a (the red and blue curve). To further confirm that the reduced AOS threshold fluence is due to the interference effect as discussed above, we compared the enhancement for different thickness of the bottom SiO$_2$ layer, as shown in Fig. \ref{fig:fig2}c. The threshold fluence of Co/Gd oscillates in a fashion (see red dashed line) corresponding to the reciprocal of the sinusoidal behaviour presented in Fig. \ref{fig:sim}b (see red curve) and a minimum is achieved at a quarter wavelength.  It is confirmed that the threshold fluence is decreased greatly by 62.5\% 3.2 mJ/cm$^2$ to 1.2 mJ/cm$^{2}$ using the bottom reflection-enhancing dielectric coating. 

We proceed by adding an extra SiO$_2$ layer on top. The measurement results from Si/t-SiO$_2$/Co/Gd/s-SiO$_2$(x) in Fig. \ref{fig:fig2}d shows that the threshold fluence displays more than two periods of oscillation in a similar fashion as the calculation results in Fig.\ref{fig:sim}b (see black curve). More specifically, as the thickness increases from 0, the threshold fluence decreases until reaching a minimum at close to a quarter wavelength. 
These results suggest that adding a top dielectric layer with proper thickness in combination with the bottom layer can bring the threshold fluence further down. 
Here we note that the threshold fluence at 0 thickness in Fig. \ref{fig:fig2}d is a higher value compared with the result in Fig. \ref{fig:fig2}a (see red data point at Co thickness equal to 1 nm). We attribute this deviation to the growth process of RF sputtering that degrades the sample, which is confirmed by a separate control measurement on the area covered completely by the wedge shutter during growth (not show). Determining the underlying mechanism is beyond the scope of this work.

Finally, to mitigate the non-idealities incurred by s-SiO$_2$ and the influence of its fabrication process on Co/Gd for the discussion of a top dielectric layer, we present the measurement result for Si/t-SiO$_2$(100)/Co/Gd/c-SiO$_2$(100). Here the top layer is deposited by ICP-CVD instead, which is shown as the blue curve in Fig. \ref{fig:fig2}b. As can be seen, the threshold fluence of 0.6 mJ/cm$^2$ (see the blue bar in the inset of Fig. \ref{fig:fig2}b and blue data point in Fig. \ref{fig:fig2}d) is almost 80\% lower than the value (3.2 mJ/cm$^2$) for an uncoated Co/Gd as expected from the simulations shown in Fig. \ref{fig:sim}a (see gray curve for a thickness of $\sim$ 0.08 $\frac{\lambda}{2 \pi}$), which could potentially reduce the non-linear losses in photonic distribution networks by more than 25 times since the nonlinear absorption scales quadratically with light intensity.

In addition, this ultra low value corresponds to a total energy consumption of 15 fJ for a device with a dimension of 50 $\times$ 50 nm$^2$. We claim this value is the lowest threshold fluence reported so far for AOS\cite{Yang:2017aa,El-Ghazaly:2020aa,Kimel_AOS_Review2019} and two orders of magnitude more energy efficient than standard SOT/STT-MRAM technologies\cite{KevinGarello:2018a,Puebla2020:a} as well as state-of-the-art SRAM and DRAM\cite{Shao:2020a,Wang:2013aa}. A further energy reduction is expected by reducing the thickness of the Co layer following results shown in Fig. \ref{fig:fig2}a, paving ways for ultrafast energy efficient on/inter-chip memory applications.

Our results show that the energy efficiency of AOS can be tuned with enhanced light absorption.
This approach can be further extended by using multilayer dielectric stacks such as Bragg mirrors sandwiching Co/Gd with constituent dielectric materials with a higher contrast in the dielectric constant, which would lead to almost total light absorption. Additionally, we propose that the full sheet coverage of the dielectric coating can be extended to patterned structures so that the heat can be selectively deposited. Such an approach could be useful for all-optical skyrmion creation by AOS and topological phase transitions\cite{Buttner:2020aa}.

In conclusion, we have theoretically analyzed the potential of applying dielectric coatings sandwiching an optically switchable magnetic layer to enhance the light absorption and thereby optimize AOS.
We used the concept of light interference enabled by SiO$_2$ dielectric coatings to reduce the threshold fluence of AOS in Co/Gd. We experimentally verified the effect by observing interference induced oscillations of the threshold fluence with respect to the $\text{SiO}_2$ thickness, which matches well with our calculation. In particular, we achieved more than 80\% of energy reduction by sandwiching the metal stack with two dielectric layers achieving an ultralow value of 0.6 mJ/cm$^2$.

\begin{acknowledgments}
This project has received funding from the European Union’s Horizon 2020 research and innovation programme under the Marie Skłodowska-Curie grant agreement No. 860060.
This work is also part of the research program Foundation for Fundamental Research on Matter (FOM) and Gravitation program "Research Center for Integrated Nanophotonics", which are financed by Dutch Research Council (NWO). 

\end{acknowledgments}
\bibliography{aipsamp}

\begin{thebibliography}{33}%
\makeatletter
\providecommand \@ifxundefined [1]{%
 \@ifx{#1\undefined}
}%
\providecommand \@ifnum [1]{%
 \ifnum #1\expandafter \@firstoftwo
 \else \expandafter \@secondoftwo
 \fi
}%
\providecommand \@ifx [1]{%
 \ifx #1\expandafter \@firstoftwo
 \else \expandafter \@secondoftwo
 \fi
}%
\providecommand \natexlab [1]{#1}%
\providecommand \enquote  [1]{``#1''}%
\providecommand \bibnamefont  [1]{#1}%
\providecommand \bibfnamefont [1]{#1}%
\providecommand \citenamefont [1]{#1}%
\providecommand \href@noop [0]{\@secondoftwo}%
\providecommand \href [0]{\begingroup \@sanitize@url \@href}%
\providecommand \@href[1]{\@@startlink{#1}\@@href}%
\providecommand \@@href[1]{\endgroup#1\@@endlink}%
\providecommand \@sanitize@url [0]{\catcode `\\12\catcode `\$12\catcode
  `\&12\catcode `\#12\catcode `\^12\catcode `\_12\catcode `\%12\relax}%
\providecommand \@@startlink[1]{}%
\providecommand \@@endlink[0]{}%
\providecommand \url  [0]{\begingroup\@sanitize@url \@url }%
\providecommand \@url [1]{\endgroup\@href {#1}{\urlprefix }}%
\providecommand \urlprefix  [0]{URL }%
\providecommand \Eprint [0]{\href }%
\providecommand \doibase [0]{http://dx.doi.org/}%
\providecommand \selectlanguage [0]{\@gobble}%
\providecommand \bibinfo  [0]{\@secondoftwo}%
\providecommand \bibfield  [0]{\@secondoftwo}%
\providecommand \translation [1]{[#1]}%
\providecommand \BibitemOpen [0]{}%
\providecommand \bibitemStop [0]{}%
\providecommand \bibitemNoStop [0]{.\EOS\space}%
\providecommand \EOS [0]{\spacefactor3000\relax}%
\providecommand \BibitemShut  [1]{\csname bibitem#1\endcsname}%
\let\auto@bib@innerbib\@empty
\bibitem [{\citenamefont {Stanciu}\ \emph {et~al.}(2007)\citenamefont
  {Stanciu}, \citenamefont {Hansteen}, \citenamefont {Kimel}, \citenamefont
  {Kirilyuk}, \citenamefont {Tsukamoto}, \citenamefont {Itoh},\ and\
  \citenamefont {Rasing}}]{Stanciu:2007aa}%
  \BibitemOpen
  \bibfield  {author} {\bibinfo {author} {\bibfnamefont {C.~D.}\ \bibnamefont
  {Stanciu}}, \bibinfo {author} {\bibfnamefont {F.}~\bibnamefont {Hansteen}},
  \bibinfo {author} {\bibfnamefont {A.~V.}\ \bibnamefont {Kimel}}, \bibinfo
  {author} {\bibfnamefont {A.}~\bibnamefont {Kirilyuk}}, \bibinfo {author}
  {\bibfnamefont {A.}~\bibnamefont {Tsukamoto}}, \bibinfo {author}
  {\bibfnamefont {A.}~\bibnamefont {Itoh}}, \ and\ \bibinfo {author}
  {\bibfnamefont {T.}~\bibnamefont {Rasing}},\ }\bibfield  {title} {\enquote
  {\bibinfo {title} {All-optical magnetic recording with circularly polarized
  light},}\ }\href@noop {} {\bibfield  {journal} {\bibinfo  {journal} {Physical
  Review Letters}\ }\textbf {\bibinfo {volume} {99}},\ \bibinfo {pages}
  {047601--} (\bibinfo {year} {2007})}\BibitemShut {NoStop}%
\bibitem [{\citenamefont {Alexey~V.Kimel}(2019)}]{Kimel_AOS_Review2019}%
  \BibitemOpen
  \bibfield  {author} {\bibinfo {author} {\bibfnamefont {M.~L.}\ \bibnamefont
  {Alexey~V.Kimel}},\ }\bibfield  {title} {\enquote {\bibinfo {title} {Writing
  magnetic memory with ultrashort light pulses},}\ }\href@noop {} {\bibfield
  {journal} {\bibinfo  {journal} {Nature Reviews Materials}\ }\textbf {\bibinfo
  {volume} {4}},\ \bibinfo {pages} {189--200} (\bibinfo {year}
  {2019})}\BibitemShut {NoStop}%
\bibitem [{\citenamefont {El-Ghazaly}\ \emph {et~al.}(2020)\citenamefont
  {El-Ghazaly}, \citenamefont {Gorchon}, \citenamefont {Wilson}, \citenamefont
  {Pattabi},\ and\ \citenamefont {Bokor}}]{El-Ghazaly:2020aa}%
  \BibitemOpen
  \bibfield  {author} {\bibinfo {author} {\bibfnamefont {A.}~\bibnamefont
  {El-Ghazaly}}, \bibinfo {author} {\bibfnamefont {J.}~\bibnamefont {Gorchon}},
  \bibinfo {author} {\bibfnamefont {R.~B.}\ \bibnamefont {Wilson}}, \bibinfo
  {author} {\bibfnamefont {A.}~\bibnamefont {Pattabi}}, \ and\ \bibinfo
  {author} {\bibfnamefont {J.}~\bibnamefont {Bokor}},\ }\bibfield  {title}
  {\enquote {\bibinfo {title} {Progress towards ultrafast spintronics
  applications},}\ }\href@noop {} {\bibfield  {journal} {\bibinfo  {journal}
  {Journal of Magnetism and Magnetic Materials}\ }\textbf {\bibinfo {volume}
  {502}},\ \bibinfo {pages} {166478} (\bibinfo {year} {2020})}\BibitemShut
  {NoStop}%
\bibitem [{\citenamefont {Becker}\ \emph {et~al.}(2020)\citenamefont {Becker},
  \citenamefont {Kr{\"u}ckel}, \citenamefont {Thourhout},\ and\ \citenamefont
  {Heck}}]{Becker:2020aa}%
  \BibitemOpen
  \bibfield  {author} {\bibinfo {author} {\bibfnamefont {H.}~\bibnamefont
  {Becker}}, \bibinfo {author} {\bibfnamefont {C.~J.}\ \bibnamefont
  {Kr{\"u}ckel}}, \bibinfo {author} {\bibfnamefont {D.~V.}\ \bibnamefont
  {Thourhout}}, \ and\ \bibinfo {author} {\bibfnamefont {M.~J.~R.}\
  \bibnamefont {Heck}},\ }\bibfield  {title} {\enquote {\bibinfo {title}
  {Out-of-plane focusing grating couplers for silicon photonics integration
  with optical mram technology},}\ }\href@noop {} {\bibfield  {journal}
  {\bibinfo  {journal} {IEEE Journal of Selected Topics in Quantum
  Electronics}\ }\textbf {\bibinfo {volume} {26}},\ \bibinfo {pages} {1--8}
  (\bibinfo {year} {2020})}\BibitemShut {NoStop}%
\bibitem [{\citenamefont {Lalieu}, \citenamefont {Lavrijsen},\ and\
  \citenamefont {Koopmans}(2019)}]{Lalieu:2019aa}%
  \BibitemOpen
  \bibfield  {author} {\bibinfo {author} {\bibfnamefont {M.~L.~M.}\
  \bibnamefont {Lalieu}}, \bibinfo {author} {\bibfnamefont {R.}~\bibnamefont
  {Lavrijsen}}, \ and\ \bibinfo {author} {\bibfnamefont {B.}~\bibnamefont
  {Koopmans}},\ }\bibfield  {title} {\enquote {\bibinfo {title} {Integrating
  all-optical switching with spintronics},}\ }\href@noop {} {\bibfield
  {journal} {\bibinfo  {journal} {Nature Communications}\ }\textbf {\bibinfo
  {volume} {10}},\ \bibinfo {pages} {110} (\bibinfo {year} {2019})}\BibitemShut
  {NoStop}%
\bibitem [{\citenamefont {Ostler}\ \emph {et~al.}(2012)\citenamefont {Ostler},
  \citenamefont {Barker}, \citenamefont {Evans}, \citenamefont {Chantrell},
  \citenamefont {Atxitia}, \citenamefont {Chubykalo-Fesenko}, \citenamefont
  {El~Moussaoui}, \citenamefont {Le~Guyader}, \citenamefont {Mengotti},
  \citenamefont {Heyderman}, \citenamefont {Nolting}, \citenamefont
  {Tsukamoto}, \citenamefont {Itoh}, \citenamefont {Afanasiev}, \citenamefont
  {Ivanov}, \citenamefont {Kalashnikova}, \citenamefont {Vahaplar},
  \citenamefont {Mentink}, \citenamefont {Kirilyuk}, \citenamefont {Rasing},\
  and\ \citenamefont {Kimel}}]{Ostler:2012aa}%
  \BibitemOpen
  \bibfield  {author} {\bibinfo {author} {\bibfnamefont {T.~A.}\ \bibnamefont
  {Ostler}}, \bibinfo {author} {\bibfnamefont {J.}~\bibnamefont {Barker}},
  \bibinfo {author} {\bibfnamefont {R.~F.~L.}\ \bibnamefont {Evans}}, \bibinfo
  {author} {\bibfnamefont {R.~W.}\ \bibnamefont {Chantrell}}, \bibinfo {author}
  {\bibfnamefont {U.}~\bibnamefont {Atxitia}}, \bibinfo {author} {\bibfnamefont
  {O.}~\bibnamefont {Chubykalo-Fesenko}}, \bibinfo {author} {\bibfnamefont
  {S.}~\bibnamefont {El~Moussaoui}}, \bibinfo {author} {\bibfnamefont
  {L.}~\bibnamefont {Le~Guyader}}, \bibinfo {author} {\bibfnamefont
  {E.}~\bibnamefont {Mengotti}}, \bibinfo {author} {\bibfnamefont {L.~J.}\
  \bibnamefont {Heyderman}}, \bibinfo {author} {\bibfnamefont {F.}~\bibnamefont
  {Nolting}}, \bibinfo {author} {\bibfnamefont {A.}~\bibnamefont {Tsukamoto}},
  \bibinfo {author} {\bibfnamefont {A.}~\bibnamefont {Itoh}}, \bibinfo {author}
  {\bibfnamefont {D.}~\bibnamefont {Afanasiev}}, \bibinfo {author}
  {\bibfnamefont {B.~A.}\ \bibnamefont {Ivanov}}, \bibinfo {author}
  {\bibfnamefont {A.~M.}\ \bibnamefont {Kalashnikova}}, \bibinfo {author}
  {\bibfnamefont {K.}~\bibnamefont {Vahaplar}}, \bibinfo {author}
  {\bibfnamefont {J.}~\bibnamefont {Mentink}}, \bibinfo {author} {\bibfnamefont
  {A.}~\bibnamefont {Kirilyuk}}, \bibinfo {author} {\bibfnamefont
  {T.}~\bibnamefont {Rasing}}, \ and\ \bibinfo {author} {\bibfnamefont {A.~V.}\
  \bibnamefont {Kimel}},\ }\bibfield  {title} {\enquote {\bibinfo {title}
  {Ultrafast heating as a sufficient stimulus for magnetization reversal in a
  ferrimagnet},}\ }\href@noop {} {\bibfield  {journal} {\bibinfo  {journal}
  {Nature Communications}\ }\textbf {\bibinfo {volume} {3}},\ \bibinfo {pages}
  {666} (\bibinfo {year} {2012})}\BibitemShut {NoStop}%
\bibitem [{\citenamefont {Lalieu}\ \emph {et~al.}(2017)\citenamefont {Lalieu},
  \citenamefont {Peeters}, \citenamefont {Haenen}, \citenamefont {Lavrijsen},\
  and\ \citenamefont {Koopmans}}]{Lalieu:2017aa}%
  \BibitemOpen
  \bibfield  {author} {\bibinfo {author} {\bibfnamefont {M.~L.~M.}\
  \bibnamefont {Lalieu}}, \bibinfo {author} {\bibfnamefont {M.~J.~G.}\
  \bibnamefont {Peeters}}, \bibinfo {author} {\bibfnamefont {S.~R.~R.}\
  \bibnamefont {Haenen}}, \bibinfo {author} {\bibfnamefont {R.}~\bibnamefont
  {Lavrijsen}}, \ and\ \bibinfo {author} {\bibfnamefont {B.}~\bibnamefont
  {Koopmans}},\ }\bibfield  {title} {\enquote {\bibinfo {title} {Deterministic
  all-optical switching of synthetic ferrimagnets using single femtosecond
  laser pulses},}\ }\href@noop {} {\bibfield  {journal} {\bibinfo  {journal}
  {Physical Review B}\ }\textbf {\bibinfo {volume} {96}},\ \bibinfo {pages}
  {220411--} (\bibinfo {year} {2017})}\BibitemShut {NoStop}%
\bibitem [{\citenamefont {Avil{\'e}s-F{\'e}lix}\ \emph
  {et~al.}(2019)\citenamefont {Avil{\'e}s-F{\'e}lix}, \citenamefont
  {{\'A}lvaro-G{\'o}mez}, \citenamefont {Li}, \citenamefont {Davies},
  \citenamefont {Olivier}, \citenamefont {Rubio-Roy}, \citenamefont {Auffret},
  \citenamefont {Kirilyuk}, \citenamefont {Kimel}, \citenamefont {Rasing},
  \citenamefont {Buda-Prejbeanu}, \citenamefont {Sousa}, \citenamefont
  {Dieny},\ and\ \citenamefont {Prejbeanu}}]{Aviles-Felix:2019aa}%
  \BibitemOpen
  \bibfield  {author} {\bibinfo {author} {\bibfnamefont {L.}~\bibnamefont
  {Avil{\'e}s-F{\'e}lix}}, \bibinfo {author} {\bibfnamefont {L.}~\bibnamefont
  {{\'A}lvaro-G{\'o}mez}}, \bibinfo {author} {\bibfnamefont {G.}~\bibnamefont
  {Li}}, \bibinfo {author} {\bibfnamefont {C.~S.}\ \bibnamefont {Davies}},
  \bibinfo {author} {\bibfnamefont {A.}~\bibnamefont {Olivier}}, \bibinfo
  {author} {\bibfnamefont {M.}~\bibnamefont {Rubio-Roy}}, \bibinfo {author}
  {\bibfnamefont {S.}~\bibnamefont {Auffret}}, \bibinfo {author} {\bibfnamefont
  {A.}~\bibnamefont {Kirilyuk}}, \bibinfo {author} {\bibfnamefont {A.~V.}\
  \bibnamefont {Kimel}}, \bibinfo {author} {\bibfnamefont {T.}~\bibnamefont
  {Rasing}}, \bibinfo {author} {\bibfnamefont {L.~D.}\ \bibnamefont
  {Buda-Prejbeanu}}, \bibinfo {author} {\bibfnamefont {R.~C.}\ \bibnamefont
  {Sousa}}, \bibinfo {author} {\bibfnamefont {B.}~\bibnamefont {Dieny}}, \ and\
  \bibinfo {author} {\bibfnamefont {I.~L.}\ \bibnamefont {Prejbeanu}},\
  }\bibfield  {title} {\enquote {\bibinfo {title} {Integration of {Tb/Co}
  multilayers within optically switchable perpendicular magnetic tunnel
  junctions},}\ }\href@noop {} {\bibfield  {journal} {\bibinfo  {journal} {AIP
  Advances}\ }\textbf {\bibinfo {volume} {9}},\ \bibinfo {pages} {125328}
  (\bibinfo {year} {2019})}\BibitemShut {NoStop}%
\bibitem [{\citenamefont {Beens}\ \emph {et~al.}(2019)\citenamefont {Beens},
  \citenamefont {Lalieu}, \citenamefont {Deenen}, \citenamefont {Duine},\ and\
  \citenamefont {Koopmans}}]{Beens:2019aa}%
  \BibitemOpen
  \bibfield  {author} {\bibinfo {author} {\bibfnamefont {M.}~\bibnamefont
  {Beens}}, \bibinfo {author} {\bibfnamefont {M.~L.~M.}\ \bibnamefont
  {Lalieu}}, \bibinfo {author} {\bibfnamefont {A.~J.~M.}\ \bibnamefont
  {Deenen}}, \bibinfo {author} {\bibfnamefont {R.~A.}\ \bibnamefont {Duine}}, \
  and\ \bibinfo {author} {\bibfnamefont {B.}~\bibnamefont {Koopmans}},\
  }\bibfield  {title} {\enquote {\bibinfo {title} {Comparing all-optical
  switching in synthetic-ferrimagnetic multilayers and alloys},}\ }\href@noop
  {} {\bibfield  {journal} {\bibinfo  {journal} {Physical Review B}\ }\textbf
  {\bibinfo {volume} {100}},\ \bibinfo {pages} {220409--} (\bibinfo {year}
  {2019})}\BibitemShut {NoStop}%
\bibitem [{\citenamefont {Avil{\'e}s-F{\'e}lix}\ \emph
  {et~al.}(2020)\citenamefont {Avil{\'e}s-F{\'e}lix}, \citenamefont {Olivier},
  \citenamefont {Li}, \citenamefont {Davies}, \citenamefont
  {{\'A}lvaro-G{\'o}mez}, \citenamefont {Rubio-Roy}, \citenamefont {Auffret},
  \citenamefont {Kirilyuk}, \citenamefont {Kimel}, \citenamefont {Rasing},
  \citenamefont {Buda-Prejbeanu}, \citenamefont {Sousa}, \citenamefont
  {Dieny},\ and\ \citenamefont {Prejbeanu}}]{Aviles-Felix:2020aa}%
  \BibitemOpen
  \bibfield  {author} {\bibinfo {author} {\bibfnamefont {L.}~\bibnamefont
  {Avil{\'e}s-F{\'e}lix}}, \bibinfo {author} {\bibfnamefont {A.}~\bibnamefont
  {Olivier}}, \bibinfo {author} {\bibfnamefont {G.}~\bibnamefont {Li}},
  \bibinfo {author} {\bibfnamefont {C.~S.}\ \bibnamefont {Davies}}, \bibinfo
  {author} {\bibfnamefont {L.}~\bibnamefont {{\'A}lvaro-G{\'o}mez}}, \bibinfo
  {author} {\bibfnamefont {M.}~\bibnamefont {Rubio-Roy}}, \bibinfo {author}
  {\bibfnamefont {S.}~\bibnamefont {Auffret}}, \bibinfo {author} {\bibfnamefont
  {A.}~\bibnamefont {Kirilyuk}}, \bibinfo {author} {\bibfnamefont {A.~V.}\
  \bibnamefont {Kimel}}, \bibinfo {author} {\bibfnamefont {T.}~\bibnamefont
  {Rasing}}, \bibinfo {author} {\bibfnamefont {L.~D.}\ \bibnamefont
  {Buda-Prejbeanu}}, \bibinfo {author} {\bibfnamefont {R.~C.}\ \bibnamefont
  {Sousa}}, \bibinfo {author} {\bibfnamefont {B.}~\bibnamefont {Dieny}}, \ and\
  \bibinfo {author} {\bibfnamefont {I.~L.}\ \bibnamefont {Prejbeanu}},\
  }\bibfield  {title} {\enquote {\bibinfo {title} {Single-shot all-optical
  switching of magnetization in {Tb/Co} multilayer-based electrodes},}\
  }\href@noop {} {\bibfield  {journal} {\bibinfo  {journal} {Scientific
  Reports}\ }\textbf {\bibinfo {volume} {10}},\ \bibinfo {pages} {5211}
  (\bibinfo {year} {2020})}\BibitemShut {NoStop}%
\bibitem [{\citenamefont {Khorsand}\ \emph {et~al.}(2012)\citenamefont
  {Khorsand}, \citenamefont {Savoini}, \citenamefont {Kirilyuk}, \citenamefont
  {Kimel}, \citenamefont {Tsukamoto}, \citenamefont {Itoh},\ and\ \citenamefont
  {Rasing}}]{Khorsand:2012aa}%
  \BibitemOpen
  \bibfield  {author} {\bibinfo {author} {\bibfnamefont {A.~R.}\ \bibnamefont
  {Khorsand}}, \bibinfo {author} {\bibfnamefont {M.}~\bibnamefont {Savoini}},
  \bibinfo {author} {\bibfnamefont {A.}~\bibnamefont {Kirilyuk}}, \bibinfo
  {author} {\bibfnamefont {A.~V.}\ \bibnamefont {Kimel}}, \bibinfo {author}
  {\bibfnamefont {A.}~\bibnamefont {Tsukamoto}}, \bibinfo {author}
  {\bibfnamefont {A.}~\bibnamefont {Itoh}}, \ and\ \bibinfo {author}
  {\bibfnamefont {T.}~\bibnamefont {Rasing}},\ }\bibfield  {title} {\enquote
  {\bibinfo {title} {Role of magnetic circular dichroism in all-optical
  magnetic recording},}\ }\href@noop {} {\bibfield  {journal} {\bibinfo
  {journal} {Physical Review Letters}\ }\textbf {\bibinfo {volume} {108}},\
  \bibinfo {pages} {127205--} (\bibinfo {year} {2012})}\BibitemShut {NoStop}%
\bibitem [{\citenamefont {Wang}\ \emph
  {et~al.}(2020{\natexlab{a}})\citenamefont {Wang}, \citenamefont {van Hees},
  \citenamefont {Lavrijsen}, \citenamefont {Zhao},\ and\ \citenamefont
  {Koopmans}}]{Wang:2020ab}%
  \BibitemOpen
  \bibfield  {author} {\bibinfo {author} {\bibfnamefont {L.}~\bibnamefont
  {Wang}}, \bibinfo {author} {\bibfnamefont {Y.~L.~W.}\ \bibnamefont {van
  Hees}}, \bibinfo {author} {\bibfnamefont {R.}~\bibnamefont {Lavrijsen}},
  \bibinfo {author} {\bibfnamefont {W.}~\bibnamefont {Zhao}}, \ and\ \bibinfo
  {author} {\bibfnamefont {B.}~\bibnamefont {Koopmans}},\ }\bibfield  {title}
  {\enquote {\bibinfo {title} {Enhanced all-optical switching and domain wall
  velocity in annealed synthetic-ferrimagnetic multilayers},}\ }\href {\doibase
  10.1063/5.0012269} {\bibfield  {journal} {\bibinfo  {journal} {Applied
  Physics Letters}\ }\textbf {\bibinfo {volume} {117}},\ \bibinfo {pages}
  {022408} (\bibinfo {year} {2020}{\natexlab{a}})}\BibitemShut {NoStop}%
\bibitem [{\citenamefont {Chen}\ \emph {et~al.}(2017)\citenamefont {Chen},
  \citenamefont {He}, \citenamefont {Wang},\ and\ \citenamefont
  {Li}}]{Chen:2017aa}%
  \BibitemOpen
  \bibfield  {author} {\bibinfo {author} {\bibfnamefont {J.-Y.}\ \bibnamefont
  {Chen}}, \bibinfo {author} {\bibfnamefont {L.}~\bibnamefont {He}}, \bibinfo
  {author} {\bibfnamefont {J.-P.}\ \bibnamefont {Wang}}, \ and\ \bibinfo
  {author} {\bibfnamefont {M.}~\bibnamefont {Li}},\ }\bibfield  {title}
  {\enquote {\bibinfo {title} {All-optical switching of magnetic tunnel
  junctions with single subpicosecond laser pulses},}\ }\href@noop {}
  {\bibfield  {journal} {\bibinfo  {journal} {Physical Review Applied}\
  }\textbf {\bibinfo {volume} {7}},\ \bibinfo {pages} {021001--} (\bibinfo
  {year} {2017})}\BibitemShut {NoStop}%
\bibitem [{\citenamefont {Wang}\ \emph
  {et~al.}(2020{\natexlab{b}})\citenamefont {Wang}, \citenamefont {Cheng},
  \citenamefont {Li}, \citenamefont {Liu}, \citenamefont {van Hees},
  \citenamefont {Lavrijsen}, \citenamefont {Lin}, \citenamefont {Cao},
  \citenamefont {Koopmans},\ and\ \citenamefont {Zhao}}]{wang2020picosecond}%
  \BibitemOpen
  \bibfield  {author} {\bibinfo {author} {\bibfnamefont {L.}~\bibnamefont
  {Wang}}, \bibinfo {author} {\bibfnamefont {H.}~\bibnamefont {Cheng}},
  \bibinfo {author} {\bibfnamefont {P.}~\bibnamefont {Li}}, \bibinfo {author}
  {\bibfnamefont {Y.}~\bibnamefont {Liu}}, \bibinfo {author} {\bibfnamefont
  {Y.~L.~W.}\ \bibnamefont {van Hees}}, \bibinfo {author} {\bibfnamefont
  {R.}~\bibnamefont {Lavrijsen}}, \bibinfo {author} {\bibfnamefont
  {X.}~\bibnamefont {Lin}}, \bibinfo {author} {\bibfnamefont {K.}~\bibnamefont
  {Cao}}, \bibinfo {author} {\bibfnamefont {B.}~\bibnamefont {Koopmans}}, \
  and\ \bibinfo {author} {\bibfnamefont {W.}~\bibnamefont {Zhao}},\ }\href@noop
  {} {\enquote {\bibinfo {title} {Picosecond switching of optomagnetic tunnel
  junctions},}\ } (\bibinfo {year} {2020}{\natexlab{b}}),\ \Eprint
  {http://arxiv.org/abs/2011.03612} {arXiv:2011.03612 [cond-mat.mes-hall]}
  \BibitemShut {NoStop}%
\bibitem [{\citenamefont {Bl{\"a}sing}\ \emph {et~al.}(2018)\citenamefont
  {Bl{\"a}sing}, \citenamefont {Ma}, \citenamefont {Yang}, \citenamefont
  {Garg}, \citenamefont {Dejene}, \citenamefont {N'Diaye}, \citenamefont
  {Chen}, \citenamefont {Liu},\ and\ \citenamefont {Parkin}}]{Blasing:2018aa}%
  \BibitemOpen
  \bibfield  {author} {\bibinfo {author} {\bibfnamefont {R.}~\bibnamefont
  {Bl{\"a}sing}}, \bibinfo {author} {\bibfnamefont {T.}~\bibnamefont {Ma}},
  \bibinfo {author} {\bibfnamefont {S.-H.}\ \bibnamefont {Yang}}, \bibinfo
  {author} {\bibfnamefont {C.}~\bibnamefont {Garg}}, \bibinfo {author}
  {\bibfnamefont {F.~K.}\ \bibnamefont {Dejene}}, \bibinfo {author}
  {\bibfnamefont {A.~T.}\ \bibnamefont {N'Diaye}}, \bibinfo {author}
  {\bibfnamefont {G.}~\bibnamefont {Chen}}, \bibinfo {author} {\bibfnamefont
  {K.}~\bibnamefont {Liu}}, \ and\ \bibinfo {author} {\bibfnamefont {S.~S.~P.}\
  \bibnamefont {Parkin}},\ }\bibfield  {title} {\enquote {\bibinfo {title}
  {Exchange coupling torque in ferrimagnetic {Co/Gd} bilayer maximized near
  angular momentum compensation temperature},}\ }\href@noop {} {\bibfield
  {journal} {\bibinfo  {journal} {Nature Communications}\ }\textbf {\bibinfo
  {volume} {9}},\ \bibinfo {pages} {4984} (\bibinfo {year} {2018})}\BibitemShut
  {NoStop}%
\bibitem [{\citenamefont {Gorchon}\ \emph {et~al.}(2016)\citenamefont
  {Gorchon}, \citenamefont {Wilson}, \citenamefont {Yang}, \citenamefont
  {Pattabi}, \citenamefont {Chen}, \citenamefont {He}, \citenamefont {Wang},
  \citenamefont {Li},\ and\ \citenamefont {Bokor}}]{Gorchon:2016aa}%
  \BibitemOpen
  \bibfield  {author} {\bibinfo {author} {\bibfnamefont {J.}~\bibnamefont
  {Gorchon}}, \bibinfo {author} {\bibfnamefont {R.~B.}\ \bibnamefont {Wilson}},
  \bibinfo {author} {\bibfnamefont {Y.}~\bibnamefont {Yang}}, \bibinfo {author}
  {\bibfnamefont {A.}~\bibnamefont {Pattabi}}, \bibinfo {author} {\bibfnamefont
  {J.~Y.}\ \bibnamefont {Chen}}, \bibinfo {author} {\bibfnamefont
  {L.}~\bibnamefont {He}}, \bibinfo {author} {\bibfnamefont {J.~P.}\
  \bibnamefont {Wang}}, \bibinfo {author} {\bibfnamefont {M.}~\bibnamefont
  {Li}}, \ and\ \bibinfo {author} {\bibfnamefont {J.}~\bibnamefont {Bokor}},\
  }\bibfield  {title} {\enquote {\bibinfo {title} {Role of electron and phonon
  temperatures in the helicity-independent all-optical switching of gdfeco},}\
  }\href@noop {} {\bibfield  {journal} {\bibinfo  {journal} {Physical Review
  B}\ }\textbf {\bibinfo {volume} {94}},\ \bibinfo {pages} {184406--} (\bibinfo
  {year} {2016})}\BibitemShut {NoStop}%
\bibitem [{\citenamefont {Latkowski}\ \emph {et~al.}(2015)\citenamefont
  {Latkowski}, \citenamefont {Moskalenko}, \citenamefont {Tahvili},
  \citenamefont {Augustin}, \citenamefont {Smit}, \citenamefont {Williams},\
  and\ \citenamefont {Bente}}]{Latkowski:2015uw}%
  \BibitemOpen
  \bibfield  {author} {\bibinfo {author} {\bibfnamefont {S.}~\bibnamefont
  {Latkowski}}, \bibinfo {author} {\bibfnamefont {V.}~\bibnamefont
  {Moskalenko}}, \bibinfo {author} {\bibfnamefont {S.}~\bibnamefont {Tahvili}},
  \bibinfo {author} {\bibfnamefont {L.}~\bibnamefont {Augustin}}, \bibinfo
  {author} {\bibfnamefont {M.}~\bibnamefont {Smit}}, \bibinfo {author}
  {\bibfnamefont {K.}~\bibnamefont {Williams}}, \ and\ \bibinfo {author}
  {\bibfnamefont {E.}~\bibnamefont {Bente}},\ }\bibfield  {title} {\enquote
  {\bibinfo {title} {Monolithically integrated 2.5 ghz extended cavity
  mode-locked ring laser with intracavity phase modulators},}\ }\href@noop {}
  {\bibfield  {journal} {\bibinfo  {journal} {Optics Letters}\ }\textbf
  {\bibinfo {volume} {40}},\ \bibinfo {pages} {77--80} (\bibinfo {year}
  {2015})}\BibitemShut {NoStop}%
\bibitem [{\citenamefont {Gonzalez}\ \emph {et~al.}(2009)\citenamefont
  {Gonzalez}, \citenamefont {Murray}, \citenamefont {Krishnamurthy},\ and\
  \citenamefont {Guha}}]{Gonzalez:2009ut}%
  \BibitemOpen
  \bibfield  {author} {\bibinfo {author} {\bibfnamefont {L.~P.}\ \bibnamefont
  {Gonzalez}}, \bibinfo {author} {\bibfnamefont {J.~M.}\ \bibnamefont
  {Murray}}, \bibinfo {author} {\bibfnamefont {S.}~\bibnamefont
  {Krishnamurthy}}, \ and\ \bibinfo {author} {\bibfnamefont {S.}~\bibnamefont
  {Guha}},\ }\bibfield  {title} {\enquote {\bibinfo {title} {Wavelength
  dependence of two photon and free carrier absorptions in inp},}\ }\href@noop
  {} {\bibfield  {journal} {\bibinfo  {journal} {Optics Express}\ }\textbf
  {\bibinfo {volume} {17}},\ \bibinfo {pages} {8741--8748} (\bibinfo {year}
  {2009})}\BibitemShut {NoStop}%
\bibitem [{\citenamefont {Marculescu}\ \emph {et~al.}(2017)\citenamefont
  {Marculescu}, \citenamefont {D{\'u}ill}, \citenamefont {Koos}, \citenamefont
  {Freude},\ and\ \citenamefont {Leuthold}}]{Marculescu:2017vj}%
  \BibitemOpen
  \bibfield  {author} {\bibinfo {author} {\bibfnamefont {A.}~\bibnamefont
  {Marculescu}}, \bibinfo {author} {\bibfnamefont {S.~{\'O}.}\ \bibnamefont
  {D{\'u}ill}}, \bibinfo {author} {\bibfnamefont {C.}~\bibnamefont {Koos}},
  \bibinfo {author} {\bibfnamefont {W.}~\bibnamefont {Freude}}, \ and\ \bibinfo
  {author} {\bibfnamefont {J.}~\bibnamefont {Leuthold}},\ }\bibfield  {title}
  {\enquote {\bibinfo {title} {Spectral signature of nonlinear effects in
  semiconductor optical amplifiers},}\ }\href@noop {} {\bibfield  {journal}
  {\bibinfo  {journal} {Optics Express}\ }\textbf {\bibinfo {volume} {25}},\
  \bibinfo {pages} {29526--29559} (\bibinfo {year} {2017})}\BibitemShut
  {NoStop}%
\bibitem [{\citenamefont {Sobolewska}\ \emph {et~al.}(2020)\citenamefont
  {Sobolewska}, \citenamefont {Pelloux-Prayer}, \citenamefont {Becker},
  \citenamefont {Li}, \citenamefont {Davies}, \citenamefont {Kr{\"u}ckel},
  \citenamefont {F{\'e}lix}, \citenamefont {Olivier}, \citenamefont {Sousa},
  \citenamefont {Prejbeanu}, \citenamefont {Kiriliouk}, \citenamefont
  {Thourhout}, \citenamefont {Rasing}, \citenamefont {Moradi},\ and\
  \citenamefont {Heck}}]{Sobolewska:2020aa}%
  \BibitemOpen
  \bibfield  {author} {\bibinfo {author} {\bibfnamefont {E.~K.}\ \bibnamefont
  {Sobolewska}}, \bibinfo {author} {\bibfnamefont {J.}~\bibnamefont
  {Pelloux-Prayer}}, \bibinfo {author} {\bibfnamefont {H.}~\bibnamefont
  {Becker}}, \bibinfo {author} {\bibfnamefont {G.}~\bibnamefont {Li}}, \bibinfo
  {author} {\bibfnamefont {C.~S.}\ \bibnamefont {Davies}}, \bibinfo {author}
  {\bibfnamefont {C.~J.}\ \bibnamefont {Kr{\"u}ckel}}, \bibinfo {author}
  {\bibfnamefont {L.~A.}\ \bibnamefont {F{\'e}lix}}, \bibinfo {author}
  {\bibfnamefont {A.}~\bibnamefont {Olivier}}, \bibinfo {author} {\bibfnamefont
  {R.~C.}\ \bibnamefont {Sousa}}, \bibinfo {author} {\bibfnamefont {I.~L.}\
  \bibnamefont {Prejbeanu}}, \bibinfo {author} {\bibfnamefont {A.~I.}\
  \bibnamefont {Kiriliouk}}, \bibinfo {author} {\bibfnamefont {D.~V.}\
  \bibnamefont {Thourhout}}, \bibinfo {author} {\bibfnamefont {T.}~\bibnamefont
  {Rasing}}, \bibinfo {author} {\bibfnamefont {F.}~\bibnamefont {Moradi}}, \
  and\ \bibinfo {author} {\bibfnamefont {M.~J.~R.}\ \bibnamefont {Heck}},\
  }\bibfield  {title} {\enquote {\bibinfo {title} {Integration platform for
  optical switching of magnetic elements},}\ }in\ \href@noop {} {\emph
  {\bibinfo {booktitle} {Proc. SPIE}}},\ Vol.\ \bibinfo {volume} {11461}\
  (\bibinfo {year} {2020})\BibitemShut {NoStop}%
\bibitem [{\citenamefont {Radu}\ \emph {et~al.}(2011)\citenamefont {Radu},
  \citenamefont {Vahaplar}, \citenamefont {Stamm}, \citenamefont {Kachel},
  \citenamefont {Pontius}, \citenamefont {D{\"u}rr}, \citenamefont {Ostler},
  \citenamefont {Barker}, \citenamefont {Evans}, \citenamefont {Chantrell},
  \citenamefont {Tsukamoto}, \citenamefont {Itoh}, \citenamefont {Kirilyuk},
  \citenamefont {Rasing},\ and\ \citenamefont {Kimel}}]{Radu:2011aa}%
  \BibitemOpen
  \bibfield  {author} {\bibinfo {author} {\bibfnamefont {I.}~\bibnamefont
  {Radu}}, \bibinfo {author} {\bibfnamefont {K.}~\bibnamefont {Vahaplar}},
  \bibinfo {author} {\bibfnamefont {C.}~\bibnamefont {Stamm}}, \bibinfo
  {author} {\bibfnamefont {T.}~\bibnamefont {Kachel}}, \bibinfo {author}
  {\bibfnamefont {N.}~\bibnamefont {Pontius}}, \bibinfo {author} {\bibfnamefont
  {H.~A.}\ \bibnamefont {D{\"u}rr}}, \bibinfo {author} {\bibfnamefont {T.~A.}\
  \bibnamefont {Ostler}}, \bibinfo {author} {\bibfnamefont {J.}~\bibnamefont
  {Barker}}, \bibinfo {author} {\bibfnamefont {R.~F.~L.}\ \bibnamefont
  {Evans}}, \bibinfo {author} {\bibfnamefont {R.~W.}\ \bibnamefont
  {Chantrell}}, \bibinfo {author} {\bibfnamefont {A.}~\bibnamefont
  {Tsukamoto}}, \bibinfo {author} {\bibfnamefont {A.}~\bibnamefont {Itoh}},
  \bibinfo {author} {\bibfnamefont {A.}~\bibnamefont {Kirilyuk}}, \bibinfo
  {author} {\bibfnamefont {T.}~\bibnamefont {Rasing}}, \ and\ \bibinfo {author}
  {\bibfnamefont {A.~V.}\ \bibnamefont {Kimel}},\ }\bibfield  {title} {\enquote
  {\bibinfo {title} {Transient ferromagnetic-like state mediating ultrafast
  reversal of antiferromagnetically coupled spins},}\ }\href {\doibase
  10.1038/nature09901} {\bibfield  {journal} {\bibinfo  {journal} {Nature}\
  }\textbf {\bibinfo {volume} {472}},\ \bibinfo {pages} {205--208} (\bibinfo
  {year} {2011})}\BibitemShut {NoStop}%
\bibitem [{\citenamefont {Kittel}(2004)}]{Kittel:2004aa}%
  \BibitemOpen
  \bibfield  {author} {\bibinfo {author} {\bibfnamefont {C.}~\bibnamefont
  {Kittel}},\ }\href {https://books.google.nl/books?id=kym4QgAACAAJ} {\emph
  {\bibinfo {title} {Introduction to Solid State Physics}}}\ (\bibinfo
  {publisher} {Wiley},\ \bibinfo {year} {2004})\BibitemShut {NoStop}%
\bibitem [{\citenamefont {Saleh}\ and\ \citenamefont
  {Teich}(2019)}]{saleh2019fundamentals}%
  \BibitemOpen
  \bibfield  {author} {\bibinfo {author} {\bibfnamefont {B.}~\bibnamefont
  {Saleh}}\ and\ \bibinfo {author} {\bibfnamefont {M.}~\bibnamefont {Teich}},\
  }\href {https://books.google.nl/books?id=rcqKDwAAQBAJ} {\emph {\bibinfo
  {title} {Fundamentals of Photonics}}},\ Wiley Series in Pure and Applied
  Optics\ (\bibinfo  {publisher} {Wiley},\ \bibinfo {year} {2019})\BibitemShut
  {NoStop}%
\bibitem [{\citenamefont {Byrnes}(2020)}]{byrnes2016multilayer}%
  \BibitemOpen
  \bibfield  {author} {\bibinfo {author} {\bibfnamefont {S.~J.}\ \bibnamefont
  {Byrnes}},\ }\bibfield  {title} {\enquote {\bibinfo {title} {Multilayer
  optical calculations},}\ }\href@noop {} {\  (\bibinfo {year} {2020})},\
  \Eprint {http://arxiv.org/abs/1603.02720} {arXiv:1603.02720
  [physics.comp-ph]} \BibitemShut {NoStop}%
\bibitem [{\citenamefont {Palik}(1985)}]{Edward-D.-Palik:1985aa}%
  \BibitemOpen
  \bibfield  {author} {\bibinfo {author} {\bibfnamefont {E.~D.}\ \bibnamefont
  {Palik}},\ }\href {\doibase
  https://doi.org/10.1016/B978-0-08-054721-3.50005-8} {\emph {\bibinfo {title}
  {Handbook of Optical Constants of Solids}}},\ edited by\ \bibinfo {editor}
  {\bibfnamefont {E.~D.}\ \bibnamefont {Palik}}\ (\bibinfo  {publisher}
  {Academic Press},\ \bibinfo {address} {Boston},\ \bibinfo {year}
  {1985})\BibitemShut {NoStop}%
\bibitem [{\citenamefont {Hacker}, \citenamefont {Katenkamp},\ and\
  \citenamefont {Fischer}(1982)}]{Hacker:1982aa}%
  \BibitemOpen
  \bibfield  {author} {\bibinfo {author} {\bibfnamefont {E.}~\bibnamefont
  {Hacker}}, \bibinfo {author} {\bibfnamefont {U.}~\bibnamefont {Katenkamp}}, \
  and\ \bibinfo {author} {\bibfnamefont {H.}~\bibnamefont {Fischer}},\
  }\bibfield  {title} {\enquote {\bibinfo {title} {{RF}-sputtered {SiO}$_2$
  films for optical applications},}\ }\href@noop {} {\bibfield  {journal}
  {\bibinfo  {journal} {Thin Solid Films}\ }\textbf {\bibinfo {volume} {97}},\
  \bibinfo {pages} {145--152} (\bibinfo {year} {1982})}\BibitemShut {NoStop}%
\bibitem [{\citenamefont {Khodier}\ and\ \citenamefont
  {Sidki}(2001)}]{Khodier:2001aa}%
  \BibitemOpen
  \bibfield  {author} {\bibinfo {author} {\bibfnamefont {S.~A.}\ \bibnamefont
  {Khodier}}\ and\ \bibinfo {author} {\bibfnamefont {H.~M.}\ \bibnamefont
  {Sidki}},\ }\bibfield  {title} {\enquote {\bibinfo {title} {The effect of the
  deposition method on the optical properties of {SiO}$_2$ thin films},}\
  }\href@noop {} {\bibfield  {journal} {\bibinfo  {journal} {Journal of
  Materials Science: Materials in Electronics}\ }\textbf {\bibinfo {volume}
  {12}},\ \bibinfo {pages} {107--109} (\bibinfo {year} {2001})}\BibitemShut
  {NoStop}%
\bibitem [{\citenamefont {Yang}\ \emph {et~al.}(2017)\citenamefont {Yang},
  \citenamefont {Wilson}, \citenamefont {Gorchon}, \citenamefont {Lambert},
  \citenamefont {Salahuddin},\ and\ \citenamefont {Bokor}}]{Yang:2017aa}%
  \BibitemOpen
  \bibfield  {author} {\bibinfo {author} {\bibfnamefont {Y.}~\bibnamefont
  {Yang}}, \bibinfo {author} {\bibfnamefont {R.~B.}\ \bibnamefont {Wilson}},
  \bibinfo {author} {\bibfnamefont {J.}~\bibnamefont {Gorchon}}, \bibinfo
  {author} {\bibfnamefont {C.-H.}\ \bibnamefont {Lambert}}, \bibinfo {author}
  {\bibfnamefont {S.}~\bibnamefont {Salahuddin}}, \ and\ \bibinfo {author}
  {\bibfnamefont {J.}~\bibnamefont {Bokor}},\ }\bibfield  {title} {\enquote
  {\bibinfo {title} {Ultrafast magnetization reversal by picosecond electrical
  pulses},}\ }\href@noop {} {\bibfield  {journal} {\bibinfo  {journal} {Science
  Advances}\ }\textbf {\bibinfo {volume} {3}},\ \bibinfo {pages} {e1603117}
  (\bibinfo {year} {2017})}\BibitemShut {NoStop}%
\bibitem [{\citenamefont {Garello}\ \emph {et~al.}(2018)\citenamefont
  {Garello}, \citenamefont {Yasin}, \citenamefont {Couet}, \citenamefont
  {Souriau}, \citenamefont {Swerts}, \citenamefont {Rao}, \citenamefont
  {Van~Beek}, \citenamefont {Kim}, \citenamefont {Liu}, \citenamefont {Kundu},
  \citenamefont {Tsvetanova}, \citenamefont {Croes}, \citenamefont {Jossart},
  \citenamefont {Grimaldi}, \citenamefont {Baumgartner}, \citenamefont
  {Crotti}, \citenamefont {Fumemont}, \citenamefont {Gambardella},\ and\
  \citenamefont {Kar}}]{KevinGarello:2018a}%
  \BibitemOpen
  \bibfield  {author} {\bibinfo {author} {\bibfnamefont {K.}~\bibnamefont
  {Garello}}, \bibinfo {author} {\bibfnamefont {F.}~\bibnamefont {Yasin}},
  \bibinfo {author} {\bibfnamefont {S.}~\bibnamefont {Couet}}, \bibinfo
  {author} {\bibfnamefont {L.}~\bibnamefont {Souriau}}, \bibinfo {author}
  {\bibfnamefont {J.}~\bibnamefont {Swerts}}, \bibinfo {author} {\bibfnamefont
  {S.}~\bibnamefont {Rao}}, \bibinfo {author} {\bibfnamefont {S.}~\bibnamefont
  {Van~Beek}}, \bibinfo {author} {\bibfnamefont {W.}~\bibnamefont {Kim}},
  \bibinfo {author} {\bibfnamefont {E.}~\bibnamefont {Liu}}, \bibinfo {author}
  {\bibfnamefont {S.}~\bibnamefont {Kundu}}, \bibinfo {author} {\bibfnamefont
  {D.}~\bibnamefont {Tsvetanova}}, \bibinfo {author} {\bibfnamefont
  {K.}~\bibnamefont {Croes}}, \bibinfo {author} {\bibfnamefont
  {N.}~\bibnamefont {Jossart}}, \bibinfo {author} {\bibfnamefont
  {E.}~\bibnamefont {Grimaldi}}, \bibinfo {author} {\bibfnamefont
  {M.}~\bibnamefont {Baumgartner}}, \bibinfo {author} {\bibfnamefont
  {D.}~\bibnamefont {Crotti}}, \bibinfo {author} {\bibfnamefont
  {A.}~\bibnamefont {Fumemont}}, \bibinfo {author} {\bibfnamefont
  {P.}~\bibnamefont {Gambardella}}, \ and\ \bibinfo {author} {\bibfnamefont
  {G.}~\bibnamefont {Kar}},\ }\bibfield  {title} {\enquote {\bibinfo {title}
  {{SOT-MRAM 300MM Integration for Low Power and Ultrafast Embedded
  Memories}},}\ \ }(\bibinfo {year} {2018})\ pp.\ \bibinfo {pages}
  {81--82}\BibitemShut {NoStop}%
\bibitem [{\citenamefont {Puebla}\ \emph {et~al.}(2020)\citenamefont {Puebla},
  \citenamefont {Kim}, \citenamefont {Kondou},\ and\ \citenamefont
  {Otani}}]{Puebla2020:a}%
  \BibitemOpen
  \bibfield  {author} {\bibinfo {author} {\bibfnamefont {J.}~\bibnamefont
  {Puebla}}, \bibinfo {author} {\bibfnamefont {J.}~\bibnamefont {Kim}},
  \bibinfo {author} {\bibfnamefont {K.}~\bibnamefont {Kondou}}, \ and\ \bibinfo
  {author} {\bibfnamefont {Y.}~\bibnamefont {Otani}},\ }\bibfield  {title}
  {\enquote {\bibinfo {title} {Spintronic devices for energy-efficient data
  storage and energy harvesting},}\ }\href {\doibase 10.1038/s43246-020-0022-5}
  {\bibfield  {journal} {\bibinfo  {journal} {Communications Materials}\
  }\textbf {\bibinfo {volume} {1}},\ \bibinfo {pages} {24} (\bibinfo {year}
  {2020})}\BibitemShut {NoStop}%
\bibitem [{\citenamefont {Shao}\ \emph {et~al.}(2021)\citenamefont {Shao},
  \citenamefont {Li}, \citenamefont {Liu}, \citenamefont {Yang}, \citenamefont
  {Fukami}, \citenamefont {Razavi}, \citenamefont {Wu}, \citenamefont {Wang},
  \citenamefont {Freimuth}, \citenamefont {Mokrousov}, \citenamefont {Stiles},
  \citenamefont {Emori}, \citenamefont {Hoffmann}, \citenamefont {Åkerman},
  \citenamefont {Roy}, \citenamefont {Wang}, \citenamefont {Yang},
  \citenamefont {Garello},\ and\ \citenamefont {Zhang}}]{Shao:2020a}%
  \BibitemOpen
  \bibfield  {author} {\bibinfo {author} {\bibfnamefont {Q.}~\bibnamefont
  {Shao}}, \bibinfo {author} {\bibfnamefont {P.}~\bibnamefont {Li}}, \bibinfo
  {author} {\bibfnamefont {L.}~\bibnamefont {Liu}}, \bibinfo {author}
  {\bibfnamefont {H.}~\bibnamefont {Yang}}, \bibinfo {author} {\bibfnamefont
  {S.}~\bibnamefont {Fukami}}, \bibinfo {author} {\bibfnamefont
  {A.}~\bibnamefont {Razavi}}, \bibinfo {author} {\bibfnamefont
  {H.}~\bibnamefont {Wu}}, \bibinfo {author} {\bibfnamefont {K.}~\bibnamefont
  {Wang}}, \bibinfo {author} {\bibfnamefont {F.}~\bibnamefont {Freimuth}},
  \bibinfo {author} {\bibfnamefont {Y.}~\bibnamefont {Mokrousov}}, \bibinfo
  {author} {\bibfnamefont {M.~D.}\ \bibnamefont {Stiles}}, \bibinfo {author}
  {\bibfnamefont {S.}~\bibnamefont {Emori}}, \bibinfo {author} {\bibfnamefont
  {A.}~\bibnamefont {Hoffmann}}, \bibinfo {author} {\bibfnamefont
  {J.}~\bibnamefont {Åkerman}}, \bibinfo {author} {\bibfnamefont
  {K.}~\bibnamefont {Roy}}, \bibinfo {author} {\bibfnamefont {J.-P.}\
  \bibnamefont {Wang}}, \bibinfo {author} {\bibfnamefont {S.-H.}\ \bibnamefont
  {Yang}}, \bibinfo {author} {\bibfnamefont {K.}~\bibnamefont {Garello}}, \
  and\ \bibinfo {author} {\bibfnamefont {W.}~\bibnamefont {Zhang}},\ }\bibfield
   {title} {\enquote {\bibinfo {title} {Roadmap of spin–orbit torques},}\
  }\href {\doibase 10.1109/TMAG.2021.3078583} {\bibfield  {journal} {\bibinfo
  {journal} {IEEE Transactions on Magnetics}\ }\textbf {\bibinfo {volume}
  {57}},\ \bibinfo {pages} {1--39} (\bibinfo {year} {2021})}\BibitemShut
  {NoStop}%
\bibitem [{\citenamefont {Wang}, \citenamefont {Alzate},\ and\ \citenamefont
  {Khalili~Amiri}(2013)}]{Wang:2013aa}%
  \BibitemOpen
  \bibfield  {author} {\bibinfo {author} {\bibfnamefont {K.~L.}\ \bibnamefont
  {Wang}}, \bibinfo {author} {\bibfnamefont {J.~G.}\ \bibnamefont {Alzate}}, \
  and\ \bibinfo {author} {\bibfnamefont {P.}~\bibnamefont {Khalili~Amiri}},\
  }\bibfield  {title} {\enquote {\bibinfo {title} {Low-power non-volatile
  spintronic memory: Stt-ram and beyond},}\ }\href@noop {} {\bibfield
  {journal} {\bibinfo  {journal} {Journal of Physics D: Applied Physics}\
  }\textbf {\bibinfo {volume} {46}},\ \bibinfo {pages} {074003} (\bibinfo
  {year} {2013})}\BibitemShut {NoStop}%
\bibitem [{\citenamefont {B{\"u}ttner}\ \emph {et~al.}(2020)\citenamefont
  {B{\"u}ttner}, \citenamefont {Pfau}, \citenamefont {B{\"o}ttcher},
  \citenamefont {Schneider}, \citenamefont {Mercurio}, \citenamefont
  {G{\"u}nther}, \citenamefont {Hessing}, \citenamefont {Klose}, \citenamefont
  {Wittmann}, \citenamefont {Gerlinger}, \citenamefont {Kern}, \citenamefont
  {Str{\"u}ber}, \citenamefont {von Korff~Schmising}, \citenamefont {Fuchs},
  \citenamefont {Engel}, \citenamefont {Churikova}, \citenamefont {Huang},
  \citenamefont {Suzuki}, \citenamefont {Lemesh}, \citenamefont {Huang},
  \citenamefont {Caretta}, \citenamefont {Weder}, \citenamefont {Gaida},
  \citenamefont {M{\"o}ller}, \citenamefont {Harvey}, \citenamefont {Zayko},
  \citenamefont {Bagschik}, \citenamefont {Carley}, \citenamefont {Mercadier},
  \citenamefont {Schlappa}, \citenamefont {Yaroslavtsev}, \citenamefont
  {Le~Guyarder}, \citenamefont {Gerasimova}, \citenamefont {Scherz},
  \citenamefont {Deiter}, \citenamefont {Gort}, \citenamefont {Hickin},
  \citenamefont {Zhu}, \citenamefont {Turcato}, \citenamefont {Lomidze},
  \citenamefont {Erdinger}, \citenamefont {Castoldi}, \citenamefont
  {Maffessanti}, \citenamefont {Porro}, \citenamefont {Samartsev},
  \citenamefont {Sinova}, \citenamefont {Ropers}, \citenamefont {Mentink},
  \citenamefont {Dup{\'e}}, \citenamefont {Beach},\ and\ \citenamefont
  {Eisebitt}}]{Buttner:2020aa}%
  \BibitemOpen
  \bibfield  {author} {\bibinfo {author} {\bibfnamefont {F.}~\bibnamefont
  {B{\"u}ttner}}, \bibinfo {author} {\bibfnamefont {B.}~\bibnamefont {Pfau}},
  \bibinfo {author} {\bibfnamefont {M.}~\bibnamefont {B{\"o}ttcher}}, \bibinfo
  {author} {\bibfnamefont {M.}~\bibnamefont {Schneider}}, \bibinfo {author}
  {\bibfnamefont {G.}~\bibnamefont {Mercurio}}, \bibinfo {author}
  {\bibfnamefont {C.~M.}\ \bibnamefont {G{\"u}nther}}, \bibinfo {author}
  {\bibfnamefont {P.}~\bibnamefont {Hessing}}, \bibinfo {author} {\bibfnamefont
  {C.}~\bibnamefont {Klose}}, \bibinfo {author} {\bibfnamefont
  {A.}~\bibnamefont {Wittmann}}, \bibinfo {author} {\bibfnamefont
  {K.}~\bibnamefont {Gerlinger}}, \bibinfo {author} {\bibfnamefont {L.-M.}\
  \bibnamefont {Kern}}, \bibinfo {author} {\bibfnamefont {C.}~\bibnamefont
  {Str{\"u}ber}}, \bibinfo {author} {\bibfnamefont {C.}~\bibnamefont {von
  Korff~Schmising}}, \bibinfo {author} {\bibfnamefont {J.}~\bibnamefont
  {Fuchs}}, \bibinfo {author} {\bibfnamefont {D.}~\bibnamefont {Engel}},
  \bibinfo {author} {\bibfnamefont {A.}~\bibnamefont {Churikova}}, \bibinfo
  {author} {\bibfnamefont {S.}~\bibnamefont {Huang}}, \bibinfo {author}
  {\bibfnamefont {D.}~\bibnamefont {Suzuki}}, \bibinfo {author} {\bibfnamefont
  {I.}~\bibnamefont {Lemesh}}, \bibinfo {author} {\bibfnamefont
  {M.}~\bibnamefont {Huang}}, \bibinfo {author} {\bibfnamefont
  {L.}~\bibnamefont {Caretta}}, \bibinfo {author} {\bibfnamefont
  {D.}~\bibnamefont {Weder}}, \bibinfo {author} {\bibfnamefont {J.~H.}\
  \bibnamefont {Gaida}}, \bibinfo {author} {\bibfnamefont {M.}~\bibnamefont
  {M{\"o}ller}}, \bibinfo {author} {\bibfnamefont {T.~R.}\ \bibnamefont
  {Harvey}}, \bibinfo {author} {\bibfnamefont {S.}~\bibnamefont {Zayko}},
  \bibinfo {author} {\bibfnamefont {K.}~\bibnamefont {Bagschik}}, \bibinfo
  {author} {\bibfnamefont {R.}~\bibnamefont {Carley}}, \bibinfo {author}
  {\bibfnamefont {L.}~\bibnamefont {Mercadier}}, \bibinfo {author}
  {\bibfnamefont {J.}~\bibnamefont {Schlappa}}, \bibinfo {author}
  {\bibfnamefont {A.}~\bibnamefont {Yaroslavtsev}}, \bibinfo {author}
  {\bibfnamefont {L.}~\bibnamefont {Le~Guyarder}}, \bibinfo {author}
  {\bibfnamefont {N.}~\bibnamefont {Gerasimova}}, \bibinfo {author}
  {\bibfnamefont {A.}~\bibnamefont {Scherz}}, \bibinfo {author} {\bibfnamefont
  {C.}~\bibnamefont {Deiter}}, \bibinfo {author} {\bibfnamefont
  {R.}~\bibnamefont {Gort}}, \bibinfo {author} {\bibfnamefont {D.}~\bibnamefont
  {Hickin}}, \bibinfo {author} {\bibfnamefont {J.}~\bibnamefont {Zhu}},
  \bibinfo {author} {\bibfnamefont {M.}~\bibnamefont {Turcato}}, \bibinfo
  {author} {\bibfnamefont {D.}~\bibnamefont {Lomidze}}, \bibinfo {author}
  {\bibfnamefont {F.}~\bibnamefont {Erdinger}}, \bibinfo {author}
  {\bibfnamefont {A.}~\bibnamefont {Castoldi}}, \bibinfo {author}
  {\bibfnamefont {S.}~\bibnamefont {Maffessanti}}, \bibinfo {author}
  {\bibfnamefont {M.}~\bibnamefont {Porro}}, \bibinfo {author} {\bibfnamefont
  {A.}~\bibnamefont {Samartsev}}, \bibinfo {author} {\bibfnamefont
  {J.}~\bibnamefont {Sinova}}, \bibinfo {author} {\bibfnamefont
  {C.}~\bibnamefont {Ropers}}, \bibinfo {author} {\bibfnamefont {J.~H.}\
  \bibnamefont {Mentink}}, \bibinfo {author} {\bibfnamefont {B.}~\bibnamefont
  {Dup{\'e}}}, \bibinfo {author} {\bibfnamefont {G.~S.~D.}\ \bibnamefont
  {Beach}}, \ and\ \bibinfo {author} {\bibfnamefont {S.}~\bibnamefont
  {Eisebitt}},\ }\bibfield  {title} {\enquote {\bibinfo {title} {Observation of
  fluctuation-mediated picosecond nucleation of a topological phase},}\
  }\href@noop {} {\bibfield  {journal} {\bibinfo  {journal} {Nature Materials}\
  } (\bibinfo {year} {2020})}\BibitemShut {NoStop}%
\end{thebibliography}%
\end{document}